\documentstyle[psfig,graphicx]{mn}

\newif\ifAMStwofonts

\def\simlt{\lower.5ex\hbox{$\; \buildrel < \over \sim \;$}}
\def\simgt{\lower.5ex\hbox{$\; \buildrel > \over \sim \;$}}

\ifoldfss
  \ifCUPmtlplainloaded \else
    \NewTextAlphabet{textbfit} {cmbxti10} {}
    \NewTextAlphabet{textbfss} {cmssbx10} {}
    \NewMathAlphabet{mathbfit} {cmbxti10} {} 
    \NewMathAlphabet{mathbfss} {cmssbx10} {} 
  \fi
  \ifAMStwofonts
    \ifCUPmtlplainloaded \else
      \NewSymbolFont{upmath} {eurm10}
      \NewSymbolFont{AMSa} {msam10}
      \NewMathSymbol{\upi}     {0}{upmath}{19}
      \NewMathSymbol{\umu}     {0}{upmath}{16}
      \NewMathSymbol{\upartial}{0}{upmath}{40}
      \NewMathSymbol{\leqslant}{3}{AMSa}{36}
      \NewMathSymbol{\geqslant}{3}{AMSa}{3E}

    \fi
  \fi
\fi 

\ifnfssone
  \newmathalphabet{\mathit}
  \addtoversion{normal}{\mathit}{cmr}{m}{it}
  \addtoversion{bold}{\mathit}{cmr}{bx}{it}
  \newmathalphabet{\mathbfit} 
  \addtoversion{normal}{\mathbfit}{cmr}{bx}{it}
  \addtoversion{bold}{\mathbfit}{cmr}{bx}{it}
  \newmathalphabet{\mathbfss} 
  \addtoversion{normal}{\mathbfss}{cmss}{bx}{n}
  \addtoversion{bold}{\mathbfss}{cmss}{bx}{n}
  \ifAMStwofonts
    \ifCUPmtlplainloaded \else
      \UseAMStwoboldmath
      \makeatletter
      \new@mathgroup\upmath@group
      \define@mathgroup\mv@normal\upmath@group{eur}{m}{n}
      \define@mathgroup\mv@bold\upmath@group{eur}{b}{n}
      \edef\UPM{\hexnumber\upmath@group}
      \new@mathgroup\amsa@group
      \define@mathgroup\mv@normal\amsa@group{msa}{m}{n}
      \define@mathgroup\mv@bold\amsa@group{msa}{m}{n}
      \edef\AMSa{\hexnumber\amsa@group}
      \makeatother
      \mathchardef\upi="0\UPM19
      \mathchardef\umu="0\UPM16
      \mathchardef\upartial="0\UPM40
      \mathchardef\leqslant="3\AMSa36
      \mathchardef\geqslant="3\AMSa3E
    \fi
  \fi
\fi 

\ifnfsstwo
  \DeclareMathAlphabet{\mathbfit}{OT1}{cmr}{bx}{it}
  \SetMathAlphabet\mathbfit{bold}{OT1}{cmr}{bx}{it}
  \DeclareMathAlphabet{\mathbfss}{OT1}{cmss}{bx}{n}
  \SetMathAlphabet\mathbfss{bold}{OT1}{cmss}{bx}{n}
  \ifAMStwofonts
    \ifCUPmtlplainloaded \else
      \DeclareSymbolFont{UPM}{U}{eur}{m}{n}
      \SetSymbolFont{UPM}{bold}{U}{eur}{b}{n}
      \DeclareSymbolFont{AMSa}{U}{msa}{m}{n}
      \DeclareMathSymbol{\upi}{0}{UPM}{"19}
      \DeclareMathSymbol{\umu}{0}{UPM}{"16}
      \DeclareMathSymbol{\upartial}{0}{UPM}{"40}
      \DeclareMathSymbol{\leqslant}{3}{AMSa}{"36}
      \DeclareMathSymbol{\geqslant}{3}{AMSa}{"3E}
    \fi
  \fi
\fi 

\ifCUPmtlplainloaded \else
  \ifAMStwofonts \else 
    \def\upi{\pi}
    \def\umu{\mu}
    \def\upartial{\partial}
  \fi
\fi

\pagerange {1--4}

\title[Cyclical period changes in Z Cha]
{Cyclical period changes in Z Chamaeleontis 
\thanks{email: bap@astro.ufsc.br (RB); chico@das.inpe.br (FJ);
 simbiotica@starmedia.com (EO); sonja@mensa.ast.uct.ac.za (SV);
 pwoudt@artemisia.ast.uct.ac.za (PAW); msc@astro.keele.ac.uk (MSC)} }
\author[R. Baptista et~al.]
       {R. Baptista $^1$, F. Jablonski $^2$, E. Oliveira $^3$, \cr
        S. Vrielmann $^4$, P. A. Woudt $^4$ and M. S. Catal\'an $^5$ \\
       $^1$ Departamento de F\'{\i}sica, UFSC, Campus Trindade, 88040-900,
        Florian\'opolis, Brazil \\
        $^2$ Divis\~ao de Astrof\'{\i}sica, Instituto Nacional de Pesquisas
        Espaciais, S.\ J.\ dos Campos, Brazil \\
        $^3$ Departamento de Astronomia - IAG, Universidade de S\~ao Paulo, 
        S\~ao Paulo, Brazil \\
        $^4$ Department of Astronomy, University of Cape Town, Rondebosch
        7700, South Africa \\
        $^5$ Astrophysics Group, School of Chemistry and Physics, Keele
        University, Keele, Stafforshire ST5 5BG }

\date{Accepted 2002 July 19. Received 2002 June 19; in original form 2002 
      June 14}
\pubyear{2002}

\begin{document}

\maketitle

\begin{abstract}
We report the identification of cyclical changes in the orbital period of
the eclipsing dwarf nova Z~Cha. We used times of mid-eclipse collected 
from the literature and our new eclipse timings to construct an
observed-minus-calculated diagram covering 30 years of observations 
(1972-2002). The data present cyclical variations that can be fitted by 
a linear plus sinusoidal function with period $28\pm 2$ yr and 
amplitude $1.0\pm 0.2$ minute.  The statistical significance of this 
period by an F-test is larger than 99.9\%.
The derived fractional period change, $\Delta P/P= 4.4 \times 10^{-7}$, 
is comparable to that of other short-period cataclysmic variables (CVs), 
but is one order of magnitude smaller than those of the long-period CVs. 
Separate fits to the first and second half of the data lead to ephemerides
with quite different cycle periods and amplitudes, indicating that the
variation is not sinusoidal or, most probably, is not strictly periodic.
The observed cyclical period change is possibly caused by a solar-type 
magnetic activity cycle in the secondary star. 
An incremental variation in the Roche lobe of the secondary star of 
$\Delta R_{L2}/R_{L2} \simeq 1.7 \times 10^{-4}$ is required in order 
to explain both the observed period change and the modulation of the
quiescent brightness previously reported by Ak, Ozkan \& Mattei. 
\end{abstract}


\section{Introduction}

Z~Cha is a short-period ($P_{orb}= 1.78$ hr) eclipsing cataclysmic
variable (CV). In these binaries, a late-type star (the secondary) 
overfills its Roche lobe and transfers matter to a companion white dwarf
(the primary). In most CVs the donor star has lower mass than 
the accreting star. Since conservative mass transfer in such situations
would lead to an increase in the orbital separation (and therefore
the cessation of mass transfer via Roche lobe overflow), the existence 
of CVs as mass-transfer binaries implies that they must continuously 
loose angular momentum in order to sustain the mass transfer process. 
As a consequence, the binary should evolve slowly towards shorter 
orbital periods (on time scales of $10^8-10^9$~yr). 
Possible mechanisms suggested for driving the continuous angular 
momentum loss are magnetic braking via the secondary star's wind 
(for $P_{orb}>3$ hr) and gravitational radiation (for $P_{orb}< 3$ hr) 
(King 1988).
At very short periods, when the secondary star becomes fully degenerate 
($M_2\simlt 0.08\;M_\odot$), mass loss leads to an expansion of this 
star and reverses the secular trend, resulting thereafter in an increasing
orbital period. However, the predicted mass transfer rate after this 
period minimum is low (\.{M}$_2 \simeq 10^{-12} \; M_\odot\; yr^{-1}$) 
and few CVs are expected to be observed in such evolutionary stage 
(Warner 1995).

The secular evolution of the binary can in principle be detected by 
measuring the changes in the orbital period of eclipsing CVs. 
Eclipses  provide a fiducial mark in time and can usually be used to 
determine the orbital period (and its derivative) with high precision.
However, attempts to measure the long-term orbital period decrease in 
CVs have been disappointing: none of the studied stars show a positive 
detection of an orbital period decrease (e.g., Beuermann \& Pakull 1984).
Instead, most of the well observed eclipsing CVs 
\footnote{i.e., those with well-sampled observed-minus-calculated (O$-$C) 
eclipse timings diagram covering more than a decade of observations.}
show cyclical period changes 
(e.g., Bond \& Freeth 1988; Warner 1988; Robinson, Shetrone \& Africano 
1991; Baptista, Jablonski \& Steiner 1992; Echevarria \& Alvares 1993; 
Wolf et~al. 1993; Baptista et~al. 1995; Baptista, Catal\'an \& Costa 2000). 
The most promising explanation of this effect seems to be the existence 
of a solar-type (quasi- and/or multi-periodic) magnetic activity cycle 
in the secondary star modulating the radius of its Roche lobe and, via
gravitational coupling, the orbital period on time scales of the order 
of a decade (Applegate 1992; Richman, Applegate \& Patterson 1994).
The relatively large amplitude of these cyclical period changes
probably contributes to mask the low amplitude, secular period decrease.

Z~Cha seemed to be a remarkable exception in this scenario.
The eclipse timings analysis of Robinson et~al. (1995) shows a conspicuous
orbital period increase on a time scale of $P/|\hbox{\.{P}}|= 2 \times
10^7$~yr, not only at a much faster rate than predicted ($\simeq 10^9$~yr)
but also with the opposite sign to the expected period decrease.

In this Letter we report new eclipse timings of Z~Cha which indicate a
clear reversal of the period increase observed by Robinson et~al. (1995).
The revised (O$-$C) diagram shows a cyclical period change similar to
that observed in many other well studied eclipsing CVs. The observations
and data analysis are presented in section \ref{observa} and the
results are discussed and summarized in section \ref{discuss}.

\section{Observations and data analysis} \label{observa}

Time-series of high-speed differential CCD photometry of Z~Cha were 
obtained on October 1995 and April 2000 at the Laborat\'orio Nacional 
de Astrof\'{\i}sica, Brazil, and on February 2002 at the South African
Astronomical Observatory. 
The observations were performed under good sky conditions and covered 
a total of 11 eclipses while the target was in quiescence. The time
resolution ranged from 10~s to 30~s. The April 2000 run was made in 
white light (W).
A summary of these observations is given in Table~\ref{dados}.
Data reduction included bias subtraction, flat-field correction, 
cosmic rays removal, aperture photometry extraction and absolute flux 
calibration.
A more complete analysis of these data will be presented in a separate
paper. Here we will concentrate on the measurement of the mid-eclipse 
times.
%
\begin{table}
\begin{minipage}{80mm}
  \caption{Log of the observations.} \label{dados}
\begin{tabular}{@{}lccl@{}}
Date 		& Filter & Telescope & Cycles \\
1995 Oct 23 & V & 1.6m/LNA  & 130\,874, 130\,875 \\
2000 Apr 14 & W & 0.6m/LNA  & 152\,817--152\,820 \\
2002 Feb 12 & R & 1.9m/SAAO & 161\,795, 161\,797, 161\,799 \\
2002 Feb 13 & R & 1.9m/SAAO & 161\,810, 161\,812 \\
\end{tabular}
\end{minipage}
\end{table}

Mid-eclipse times were measured from the mid-ingress and mid-egress
times of the white dwarf eclipse using the derivative technique described
by Wood, Irwin \& Pringle (1985). 
For a given observational season, all light curves were phase-folded 
according to a test ephemeris and sorted in phase to produce a combined 
light curve with increased phase resolution. The combined light curve is 
smoothed with a median filter and its numerical derivative is calculated. 
A median-filtered version of the derivative curve is then analyzed by an
algorithm which identifies the points of extrema (the mid-ingress\,/\,egress
phases of the white dwarf). The mid-eclipse phase, $\phi_0$, is the mean 
of the two measured phases. 
Finally, we adopt a cycle number representative of the ensemble of light
curves and compute the corresponding observed mid-eclipse time (HJD) for
this cycle including the measured value of $\phi_0$. This yields a single, 
but robust mid-eclipse timing estimate from a sample of eclipse light 
curves. These measurements have a typical accuracy of about 4~s.

For Z~Cha the difference between universal time (UT) and terrestrial 
dynamical time (TDT) scales amounts to 19~s over the data set. 
The difference between the baricentric and the heliocentric corrections 
is smaller than 1~s as Z~Cha is close to the ecliptic pole.
The mid-eclipse timing have been calculated on the solar system baricentre
dynamical time (BJDD), according to the code by Stumpff (1980).
The terrestrial dynamical (TDT) and ephemeris (ET) time scales were assumed
to form a contiguous scale for our purposes.
The new eclipse timings are listed in Table~\ref{timings}.
The corresponding uncertainties in the last digit are indicated in 
parenthesis.
%
%
\begin{table}
\begin{minipage}{80mm}
  \caption{New eclipse timings.} \label{timings}
\begin{tabular}{@{}cccl@{}}
Cycle  & HJD & BJDD & ~(O$-$C) \footnote{Observed minus calculated times
with respect to the linear ephemeris of Table~\ref{zcha.efem}.} \\ [-0.5ex]
     & (2,450,000+) & (2,450,000+) & (cycles) \\ [1ex]
130\,874 & 0014.70294 & 0014.70358 (4) & $+0.0115$ \\
152\,818 & 1649.51443 & 1649.51515 (4) & $-0.0043$ \\
161\,803 & 2318.89047 & 2318.89119 (4) & $-0.0072$ \\ [-4ex]
\end{tabular}
\end{minipage}
\end{table}

The data points were weighted by the inverse of the squares of the
uncertainties in the mid-eclipse times. We arbitrarily adopted equal 
errors of $5\times 10^{-5}$~d for the optical timings in the literature 
(Cook \& Warner 1984; Cook 1985; Wood et~al. 1986; van Amerongen, 
Kuulkers \& van Paradijs 1990), and a smaller error of 
$2\times 10^{-5}$~d for the timing of Robinson et~al. (1995), half
the error in our measurements, and combined the 74 timings to compute
revised ephemerides for Z~Cha.
Table~\ref{zcha.efem} presents the parameters of the best-fit linear, 
quadratic and linear plus sinusoidal ephemerides with their 1-$\sigma$ 
formal errors quoted.
We also list the root-mean-squares of the residuals and the 
$\chi^{2}_{\nu_{2}}$ value for each case, where $\nu_{2}$ is the 
number of degrees of freedom.
%
%
\begin{table}
\centering
 \begin{minipage}{80mm}
 \caption{Ephemerides of Z\,Cha}  \label{zcha.efem}
\begin{tabular}{@{}ll@{}}

\multicolumn{2}{l}{\bf Linear ephemeris:} \\
\multicolumn{2}{l}{BJDD = T$_{0}$ + P$_{0}\cdot E$} \\ [1ex]
T$_{0} = 2440264.68070\,(\pm 4)$ d &
P$_{0} = 0.0744993048\,(\pm 5)$ d \\
$\chi^{2}_{\nu_{2}}= 49.2, \;\;\; \nu_{2}$ = 72 &
$\sigma_{1}= 3.27 \times 10^{-3}$ cycles \\ [2ex]

\multicolumn{2}{l}{\bf Quadratic ephemeris:} \\ 
\multicolumn{2}{l}{BJDD = T$_{0}$ + P$_{0}\cdot E$ + $c\cdot E^{2}$} \\ [1ex]
T$_{0} = 2440264.68088\,(\pm 8)$ d & 
P$_{0} = 0.074499300\,(\pm 2)$ d \\
c $ = (+2.8 \pm 1.1) \times 10^{-14}$ d &
$\sigma_{2} = 3.13 \times 10^{-3}$ cycles \\ 
$\chi^{2}_{\nu_{2}}= 48.0, \;\;\; \nu_{2}$ = 71 \\ [2ex]

\multicolumn{2}{l}{\bf Sinusoidal ephemeris:} \\
\multicolumn{2}{l}{BJDD = T$_{0}$ + P$_{0}\cdot E$ + A$\cdot 
\cos\,[2\pi (E-{\rm B})/{\rm C}]$} \\ [1ex]
T$_{0}= 2440264.6817\,(\pm 1)$ d & B $= (120 \pm 4) \times 10^{3}$ cycles \\
P$_{0}= 0.074499297\,(\pm 2)$ d  & C $= (136 \pm 7) \times 10^{3}$ cycles \\
A $= (7.2 \pm 1.0) \times 10^{-4}$ d & 
$\sigma_{\rm S}= 1.09 \times 10^{-3}$ cycles \\
$\chi^{2}_{\nu_{2}}= 3.99, \;\;\; \nu_{2}= 69$ \\
\end{tabular}
\end{minipage}
\end{table}
%

\begin{figure}	
\includegraphics[bb=1cm 1cm 20cm 26.5cm,scale=0.45]{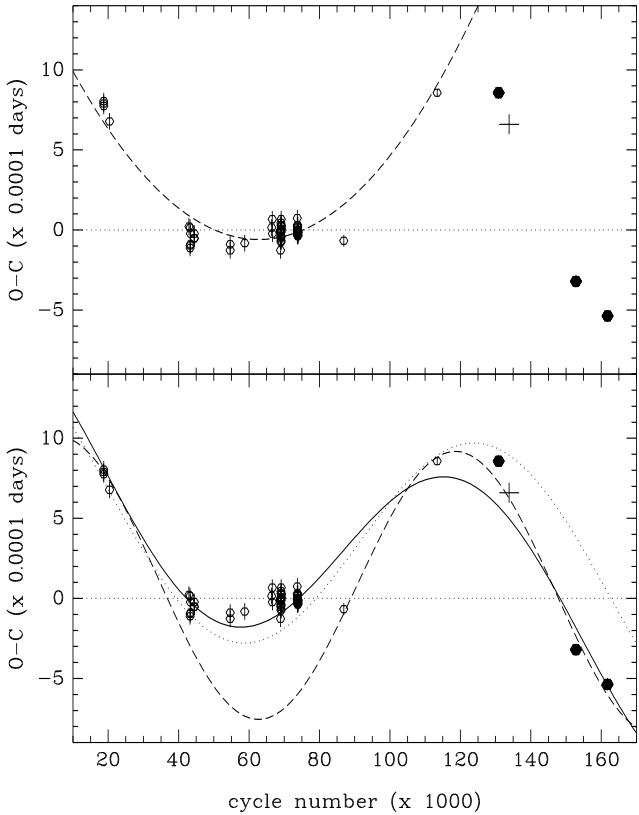}
\caption{ The (O$-$C) diagram of Z Cha with respect to the linear ephemeris
  of Table~\ref{zcha.efem}. The optical timings from the literature are 
  shown as open circles, the x-ray timing from van Teeseling (1997) is
  shown as a cross, and the new timings are indicated as solid circles.
  The dashed line in the upper panel depicts the quadratic ephemeris 
  of Robinson et~al (1995) while the solid line in the lower panel shows
  the best-fit linear plus sinusoidal ephemeris of Table~\ref{zcha.efem}. 
  Best-fit linear plus sinusoidal ephemerides for the data on the first 
  half of the time interval ($E<80\times 10^3$ cycle, dotted curve) and 
  for the last half of the time interval ($E>80 \times 10^3$ cycle, dashed
  curve) are also shown in the lower panel. }
  \label{fig1}
\end{figure}

Fig.~\ref{fig1} presents the (O$-$C) diagram with respect to the linear 
ephemeris in Table~\ref{zcha.efem}.
van Teeseling (1997) reports that a positive offset of 90~s (plus
an additional offset of 0.0025 cycles) was needed in order to centre
the x-ray eclipse of Z~Cha around phase zero with respect to the
ephemeris of Robinson et~al. (1995). We used this information to
assign a representative eclipse cycle to his measurement and to
compute the (O$-$C) value with respect to the linear ephemeris in 
Table~\ref{zcha.efem}. His x-ray timing is plotted as a cross in 
Fig.~\ref{fig1} and is fully consistent with our results.
%
%

The significance of adding additional terms to the linear ephemeris
was estimated by using the F-test, following the prescription of 
Pringle (1975).  Not surprisingly, the quadratic ephemeris is no longer
statistically significant. From the $\sigma_{1}$ and $\sigma_{\rm S}$ 
values in Table~\ref{zcha.efem} we obtain $F(3,69) = 130.7$, which 
corresponds to a 99.99\% confidence level for the linear plus sinusoidal
ephemeris with respect to the linear fit.
The best-fit linear plus sinusoidal ephemeris is shown as a solid line in 
the lower panel of Fig.~\ref{fig1}.
  
However, the eclipse timings show systematic and significant deviations 
from the best-fit linear plus sinusoidal ephemeris.  The fact that
$\chi^{2}_{\nu_{2}}>1$ emphasizes that the linear plus sinusoidal 
ephemeris is not a complete description of the data, perhaps signalling 
that the period variation is not sinusoidal or not strictly periodic.
We explored these possibilities by making separate fits to the first 
half of the data set ($E<80\times 10^3$ cycle) and to the second half
of the data set ($E>80 \times 10^3$ cycle). The results are shown in the
lower panel of Fig.~\ref{fig1}, respectively, as dotted and dashed curves.
The older timings point towards a longer cycle period with a lower 
amplitude (A= 1.13 min, P$_{cyc}= 29$~yr), whereas the more recent 
timings are best described by a shorter cycle period of larger amplitude
(A= 1.23 min, P$_{cyc}= 23$~yr).
These results indicate that the period changes are not sinusoidal or 
not periodic.

\section{Discussion} \label{discuss}

Our results reveal that the orbital period of Z~Cha is no longer increasing
as previously found by Robinson et~al. (1995). Instead, the (O$-$C) diagram
shows conspicuous cyclical, quasi-periodic changes of amplitude 1~min on a 
time-scale of about 28 yr. 

Cyclical orbital period changes are seen in many eclipsing CVs (Warner 1995
and references therein). The cycle periods range from 4~yr in EX~Dra 
(Baptista et~al. 2000) to about 30~yr in UX~UMa (Rubenstein, Patterson \& 
Africano 1991), whereas the amplitudes are in the range 0.1-2.5 min. 
Therefore, Z~Cha fits nicely in the overall picture drawn from the 
observations of orbital period changes in CVs.

If one is to seek for a common explanation for the cyclical period
changes in CVs, then models involving apsidal motion or a third body
in the system shall be discarded as these require that the orbital period 
change be strictly periodic, whereas the observations show that this
is not the case (Richman et~al. 1994 and references therein).
We may also discard explanations involving angular momentum exchange in
the binary, as cyclical exchange of rotational and orbital angular 
momentum (Smak 1972; Biermann \& Hall 1973) requires discs with masses 
far greater than those deduced by direct observations, and 
the time-scales required to allow the spin-orbit coupling of a secondary 
of variable radius are much shorter than the tidal synchronization scales 
for these systems ($\sim 10^{4}$ yr, see Applegate \& Patterson 1987).

The best current explanation for the observed cyclical period modulation 
is that it is the result of a solar-type magnetic activity cycle in the
secondary star (Applegate \& Patterson 1987; Warner 1988; Bianchini 1990). 
Richman et~al. (1994) proposed a model in which the Roche lobe radius 
of the secondary star $R_{L2}$ varies in response to changes in the 
distribution of angular momentum inside this star (caused by the magnetic
activity cycle), leading to a change in the orbital separation and, 
therefore, in the orbital period. 
As a consequence of the change in the Roche lobe radius, the mass 
transfer rate \.{M}$_2$ also changes. 
In this model, the orbital period is the shortest when the secondary star 
is the most oblate (i.e., its outer layers rotate faster), and is the 
longest when the outer layers of the secondary star are rotating the 
slowest.

The fractional period change $\Delta P/P$ is related to the amplitude
$\Delta(O-C)$ and to the period $P_{mod}$ of the modulation by 
(Applegate 1992),
\begin{equation}
\frac{\Delta P}{P}= 2\pi\; \frac{\Delta(O-C)}{P_{mod}}= 2\pi\; \frac{A}{C} \; .
\label{eq:pponto}
\end{equation}
Using the values of $A$ and $C$ in Table~\ref{zcha.efem}, we find 
$\Delta P/P = 4.4 \times 10^{-7}$.
This fractional period change is comparable to those of other short-period
CVs, but is one order of magnitude smaller than those of the CVs above 
the period gap ($\Delta P/P \simeq 2\times 10^{-6}$) [Warner 1995]. 

The predicted changes in Roche lobe radius and mass transfer rate are 
related to the fractional period change by (Richman et~al. 1994),
\begin{equation}
\frac{\Delta R_{L2}}{R_{L2}}= 39\; \left( \frac{1+q}{q} \right)^{2/3}
\left( \frac{\Delta\Omega}{10^{-3}\Omega} \right)^{-1} \frac{\Delta P}{P} \; ,
\end{equation}
and by,
\begin{equation}
\frac{\Delta \hbox{\.{M}}_2}{\hbox{\.{M}}_2} =
1.22 \times 10^5 \; \left( \frac{1+q}{q} \right)^{2/3}
\left( \frac{\Delta\Omega}{10^{-3}\Omega} \right)^{-1} \frac{\Delta P}{P} \; ,
\end{equation}
where $\Delta\Omega/\Omega$ is the fractional change in the rotation
rate of the outer shell of the secondary star involved in the cyclical 
exchange of angular momentum, $q\; (= M_2/M_1)$ is the binary mass ratio, 
and the minus signs were dropped. 

In the framework of the disc instability model (Smak 1984, Warner 1995 
and references therein), the mass transferred from the secondary star 
will accumulate in the outer disc regions during the quiescent phase. 
Hence, changes in \.{M}$_2$ will mainly affect the luminosity $L_{bs}$
of the bright spot where the infalling material hits the outer edge of 
the disc. 
The luminosity of the bright spot is $L_{bs} \propto \hbox{\.{M}}_2$. 
The bright spot in Z~Cha contributes about 30 per cent of the total 
optical light (on an average over the orbital cycle) [Wood et~al. 1986].
Therefore, one expects that the changes in mass transfer rate lead to
a modulation in the quiescent brightness of the system of $\Delta m = 
(\Delta L_{bs}/L) \simeq 0.3 \; (\Delta$\.{M}$_2$/\.{M}$_2) \simeq 
(0.06 - 0.12)$~mag for $\Delta\Omega/\Omega = (1-2)\times 10^{-3}$
(Richman et~al. 1994) and $q=0.15$ (Wood et~al. 1986).

Ak, Ozkan \& Mattei (2001) reported the detection of cyclical
modulations of the quiescent magnitudes and outburst intervals of a set
of dwarf novae, which they interpreted as the manifestation of a magnetic
activity cycle in their secondary stars. They found that the quiescent 
brightness of Z~Cha is modulated with an amplitude of $\Delta m= 0.16$~mag
on a time scale of 14.6~yr. Unfortunately, they do not list the epoch
of maximum brightness.

The observed amplitude of the brightness modulation is larger than
that predicted from the orbital period change with the assumption of
$\Delta\Omega/\Omega = (1-2)\times 10^{-3}$.
The model of Richman et~al. (1994) can account for both the measured 
period changes and the observed brightness modulation if 
$\Delta\Omega/\Omega \simeq 2.7 \times 10^{-3}$. This yields
$\Delta R_{L2}/R_{L2} \simeq 1.7 \times 10^{-4}$.

The predicted change in the Roche lobe radius of the secondary star in
Z~Cha is comparable to the observed change in the radius of the Sun as
a consequence of its magnetic activity cycle (Gilliland 1981).
The period of the quiescent brightness modulation is very close to half
of the period of the observed $P_{mod}$ and may correspond to the first 
harmonic of a non-sinusoidal or non-strictly periodic period change. 

The analysis of Ak et~al. (2001) covers only 18 years of observations. 
It would be interesting to see whether the 28~yr orbital period change
also appear as a modulation in the quiescent magnitude in a dataset 
covering a larger time-interval.
A simple and interesting test of the Richman et~al. (1994) model is to 
check its prediction that the maximum of the brightness modulation 
coincides with the minimum of the orbital period modulation, which occurred 
at $E \simeq 5.2 \times 10^4$ cycle (or about JD 2\,444\,150) for Z~Cha.

Finally, we remark that, if the magnetic activity cycle explanation is
right, our confirmation that Z~Cha also shows cyclical period changes
underscores the conclusion of Ak et~al. (2001) that even fully convective
secondary stars possess magnetic activity cycles (and, therefore, 
magnetic fields).

\section*{Acknowledgments}

We are grateful to the LNA staff for valuable assistance during the
observations.
This work was partially supported by the PRONEX/Brazil program through
the research grant FAURGS/FINEP 7697.1003.00. RB acknowledges financial 
support from CNPq through grant no. 300\,354/96-7. 
EO acknowledges financial support from FAPESP through grant no. 99/05603-6.
SV thanks the South African Claude Harris Leon Foundation for funding a
postdoctoral fellowship.

\bsp

\end{document}